\documentclass[a4paper,11pt]{article}
\usepackage{aaskaiid}
\usepackage{orcidlink}

\title{Study of Polarized Emission in  Radio Halos and Filaments in the SKA Telescopes Era}
\ShortTitle{Polarized emission in  radio halos and filaments}

\author[1,2]{Valentina Vacca\orcidlink{0000-0003-1997-0771}}
\ShortName{Vacca et al.} 
\author[1]{Federica Govoni}
\author[1]{Matteo Murgia}
\author[1]{Paolo Marchegiani\orcidlink{0000-0001-7487-8287}}
\author[4]{Hui Li}
\author[3,7]{Myriam Gitti\orcidlink{0000-0002-0843-3009}}
\author[1]{Francesca Loi}
\author[3]{Luigina Feretti}
\author[3]{Ettore Carretti}
\author[5]{Elia Battistelli}
\author[1,2,10]{Andrea Cabriolu}
\author[2,6]{Torsten A. En{\ss}lin}
\author[8]{Chiara Ferrari}
\author[3,7]{Gabriele Giovannini}
\author[9]{Richard A. Perley}

\affiliation[1]{INAF-Osservatorio Astronomico di Cagliari, Via della Scienza 5, I-09047 Selargius (CA), Italy}
\emailAdd{valentina.vacca@inaf.it}
\affiliation[2]{Max Planck Institute for Astrophysics, Karl-Schwarzschildstr. 1, 85741 Garching, Germany}
\affiliation[3]{INAF - Istituto di Radioastronomia, Via P. Gobetti 101, 40129 Bologna, Italy}
\affiliation[4]{Theoretical Astrophysics, Los Alamos National Laboratory, Los Alamos, NM, USA}
\affiliation[5]{Sapienza—University of Rome—Physics department, Piazzale Aldo Moro 5—I-00185, Rome, Italy}
\affiliation[6]{Deutsches Zentrum für Astrophysik, Postplatz 1, 02826 Görlitz, Germany}
\affiliation[7]{Dipartimento di Fisica e Astronomia – University of Bologna, via Gobetti 93/2, Italy}
\affiliation[8]{Université Côte d'Azur, OCA, CNRS, Laboratoire Lagrange, Boulevard de l'Observatoire, CS 34229, 06304 Nice Cedex 4, France}
\affiliation[9]{National Radio Astronomy Observatory, P.O.Box O, Socorro, NM, 87801}
\affiliation[10]{Department of Mathematics and Computer Science, University of Cagliari, Via Ospedale 72, Cagliari, 09121, Italy}

\abstract{Synchrotron diffuse emission in merging galaxy clusters and along filaments connecting them demonstrates the presence of relativistic particles and magnetic fields in these environments. The study of the polarized signal associated with this emission represents a powerful tool to constrain the properties of intracluster magnetic fields and the physics of acceleration and transport of relativistic particles. Despite technological progress, detecting this polarized signal is still very challenging. In order to shed light on the capabilities of the SKA telescopes to study this emission, we use the data of cosmological magneto-hydro-dynamic simulations to predict the expected polarized surface brightness of diffuse synchrotron sources from the center of galaxy clusters to filaments of the cosmic web at 1.4\,GHz. We explore the possibility to detect these sources with a polarization survey with SKA-Mid with AA4 telescopes and compare the results with those from pointed observations corresponding to longer exposure times. These simulations provide precious information to understand the potential of the SKA telescopes for studying the origin and evolution of cosmological magnetic fields.
We discuss how these observations can be used in order to characterize the magnetic field and the distribution and energy content of the radio emitting plasma and to shed light on the link between non-thermal and thermal properties and the dynamical state of the system.
}

\begin{document}
\maketitle

\section{Introduction}
Diffuse synchrotron sources called radio halos are well known to inhabit the central regions of a fraction of merging galaxy clusters: these sources are characterized by low radio brightness ($\sim$0.1\,$\mu$Jy/arcsec$^2$ at 1.4\,GHz) and a steep integrated spectrum ($S_{\nu}\propto \nu^{-\alpha}$, with $\alpha\approx1-1.3$), and are typically extended on $\gtrsim$ Mpc scales \citep{vanWeeren2019}. These sources are believed to be powered by turbulence produced in merging galaxy clusters that (re-)accelerates a pre-existing electron population embedded in a $\sim\mu$G intracluster magnetic field to ultra-relativistic energies ($\gamma\gtrsim 10^4$), see e.g. \cite{Brunetti2014}. 
\cite{Govoni2019} observed for the first time diffuse synchrotron emission in total intensity along the filament of gas \citep{Planck2013,Planck2016} connecting two galaxy clusters A399 and A401, extended on a scale of about 12\,Mpc \citep{Hincks2022}. Since then, other few cases have been found in the system A1758 \citep{Botteon2020}, along a bridge connecting a cluster and a group in the Shapley Supercluster \citep{Venturi2022}, beyond the radio halo in A2061 towards A2067 \citep{Pignataro2024}, and connecting the galaxy cluster A3017 with a potential galaxy group and extending in the direction of A3016 \citep{Hu2025}. 
 
Detecting polarized emission from these diffuse synchrotron sources is very challenging. To date filamentary structures of polarized emission associated with radio halos have been found only in three cases: A2255 
\citep{Govoni2005}, MACS\,J0717.5 +3745 \citep{Bonafede2009}, and A523 \citep{Girardi2016}. \cite{Vacca2022} found that the polarized emission in A523 is more extended ($\sim$ 2.5\,Mpc) than previously observed, and it has been 
 detected also where the total intensity counterpart is below the confusion noise. This is the first time that a polarized signal associated with the intracluster medium at the centre of a galaxy cluster is detected on such large scales, indicating the presence of non-thermal components also in the cluster outskirts. However, A523 is characterized by a low Galactic latitude (b=-20.15$^{\circ}$), therefore a possible Galactic origin of the more external patches of this emission can not be completely excluded. 
 Polarized emission from filaments of the cosmic web has been also detected, even if statistically only, through the polarised signal associated with accretion shocks \citep{Vernstrom2023}. 
Analysis of total and polarized brightness of radio halos and from filaments of the cosmic web  provides a powerful tool for studying magnetization of the environment.
 While total intensity depends on both magnetic field and relativistic electrons, the ratio between the polarized and the total intensity (fractional polarization) carries out information mainly on the magnetic field. However, the observed fractional polarization is typically lower than the intrinsic value due to the following effects.
 First, the observed polarized intensity is the result of a vectorial sum involving random orientations that reflect the tangled structure of the magnetic field in these environments. Second, the
radio signal is subject to Faraday rotation
as it passes through the intracluster magneto-ionic medium. These two effects cause the so-called internal depolarization of the signal. Furthermore, convolution with the observing beam
further suppresses the polarized signal due to small-scale fluctuations of the magnetic field and/or thermal gas density inside the resolution element. This resulting extremely faint signal, often, is completely buried
in the noise, making it very difficult to detect, despite its importance \citep[see][for further details]{Vacca2010,Govoni2013}.
Resorting to numerical simulations, \cite{Govoni2013} and \cite{Govoni2015} demonstrated that future data from the Square Kilometre Array Observatory (SKAO) should allow us to observe polarized diffuse synchrotron emission from the central regions of galaxy clusters with X-ray luminosity $\gtrsim 6\times 10^{44}$\,erg/s, thanks to enanched sensitivity and resolution. 
Recently, \cite{Vacca2024} have shown that mid-frequencies SKAO polarisation observations are not significantly affected by confusion noise and that polarized diffuse synchrotron emission from galaxy clusters and filaments connecting them can be detected with deep observations even if the total intensity signal is completely buried in the confusion noise.  

In this paper, we follow-up that work by presenting expectations of detecting polarized emission from galaxy clusters and filaments connecting them with SKA-Mid AA4 telescopes both for survey and pointed observations, considering different properties for the energy distributions of the  relativistic electrons. For expectations with SKA-Low telescopes, we refer to \cite{Vacca2024}. Here, we use a $\Lambda$CDM cosmology with $H_0$ = 72\,km/s/Mpc, $\Omega_{\rm m}$ = 0.258, and $\Omega_{\Lambda}$ = 0.742. In the following, we consider a simulated system that sits at a redshift of $z=0.073$. With the adopted cosmology, at this distance, 1$^{\prime\prime}$ corresponds to 1.354\,kpc.

A complementary approach to study magnetic field properties is represented by the analysis of the Faraday effect on background radio galaxies. We refer to the chapters by \cite{Carcamo2026}, \cite{Loi2026}, \cite{Kurahara2026}, \cite{OSullivan2026}, and \cite{Vacca2026} in this book, for the application of this technique.

\section{Simulations}

In order to assess the potential of SKA-Mid AA4 telescopes in revealing polarized emission associated with radio halos and filaments connecting galaxy clusters, we resort to a cosmological magneto-hydro-dynamic (MHD) simulation produced by the group of Hui Li at Los Alamos National Laboratories, USA. 
The simulation runs from z = 30 to z = 0, following the evolution of dark matter, baryonic
matter, and magnetic fields. It uses an adiabatic equation of state with a specific heat ratio, $\Gamma$ =5/3, and does not include heating and cooling physics or chemical reactions. Being interested in diffuse synchrotron emission that current observations suggest to be associated with turbulence caused by the merging processes \citep{Brunetti2014}, we expect that the contributions from cooling and other heating processes are negligible.
Magnetic fields are injected by active galactic nuclei at z = 2–3 and then amplified and spread over megaparsec scales during the late stages
of the merger. As pointed out by  \cite{Xu2010}, the resulting intracluster magnetic field distribution at low redshift is not deeply affected by the injection mechanism. 
For further details about the simulation, please refer to \cite{Xu2008,Xu2009,Xu2012}.
In the following, we use a single snapshot of the simulation that captures a pair
of galaxy clusters and the filament connecting them, prior the merging process begins. The output of the simulation consists of temperature, thermal plasma density, and magnetic field three-dimensional cubes with size $\approx$(6.42\,Mpc)$^3$ and resolution $\approx$(10.7\,kpc)$^3$ \footnote{This value is a factor four larger than the linear scale corresponding to the minimum angular resolution considered in this chapter, i.e. 2$^{\prime\prime}$ (2.7\,kpc at the distance of the cluster). However, we do not expect it affects our results because the thermal gas density of the system is smooth and the magnetic field fluctuates on spatial scales of the order of hundreds of kpc, preventing depolarization inside the resolution element \citep[see Fig.\,1 and Fig.\,14 respectively in][]{Vacca2024}.}.

By using the magnetic field cube,
we produced synthetic radio images of the system. 
As proposed by \cite{Murgia2004}, total intensity and intrinsic linear polarisation emissivity at each point on the computational grid have been calculated by convolving
the emission spectrum of a single relativistic electron with the
particle energy distribution of an isotropic population of relativistic electrons. In order to produce Stokes parameters Q and U images, we took into account internal depolarization due to Faraday rotation suffered by the signal while crossing the intracluster magneto-ionic medium. To this end we exploited both the magnetic field and the thermal
gas density cubes.
We consider here synthetic images of the diffuse emission only, because radio galaxy emission is beyond the purposes of this analysis and has been already addressed in \cite{Loi2019b}.

A detailed description and characterization of the intracluster medium (ICM) physical parameters in the simulation and of the parameters assumed for the relativistic particles distribution to produce synthetic radio images is given in \cite{Vacca2024}.

\section{Results}

In the following, we present synthetic radio images in total intensity and polarization for observations with SKA-Mid in Band\,2 (frequency range 0.95 - 1.76\,GHz). 
The sensitivities have been derived with the SKAO Sensitivity Calculator for mid frequencies\footnote{\url{https://sensitivity-calculator.skao.int/mid}}, by adopting uniform weighting that minimises the side lobes and yields a more Gaussian point-spread-function. Radio images have been obtained assuming two scenarios for the energy density of the relativistic electron population:
 (I) equipartition between the magnetic
field ($u_{\rm B}$) and the relativistic electron ($u_{\rm el}$) energy density at
every location in the ICM (we assume that all the magnetic energy at a given spatial location is converted in energy of emitting particles and consider a filling factor equal to 1); 
(II) a relativistic
electron distribution with an energy density equal to 0.3\% of the thermal one ($u_{\rm th}$), this factor ensures that the radio power at 1.4\,GHz is consistent with that of radio halos in equipartition conditions and
with the upper limit from $\gamma$-ray observations \citep[e.g.,][]{Loi2019b,Brunetti2017}.

The Magnetism SKAO Science Working Group identified as top priority a polarimetric survey in Band\,2, with 1\,MHz of channel width,  2$^{\prime\prime}$ of spatial resolution, and 15\,min of observing time, with the aim of achieving a sensitivity of $\sim$4\,$\mu$Jy/beam per Stokes parameter \citep{Heald2020}. Here, we assume a similar observing setup, but with a slightly reduced frequency band, i.e. between 0.95-1.67\,GHz, because of the 13.5\,m MeerKAT antennas. This translates into a sensitivity of $\sigma_{\rm I}=\sigma_{\rm Q,U}=6.0\,\mu$Jy/beam. The confusion noise is negligible both in total intensity and in polarization, being 100-150\,nJy/beam in Stokes I and about 0.4\,nJy/beam in Stokes Q and U. 
\begin{figure}[h]
    \centering
	\includegraphics[width=1\columnwidth]{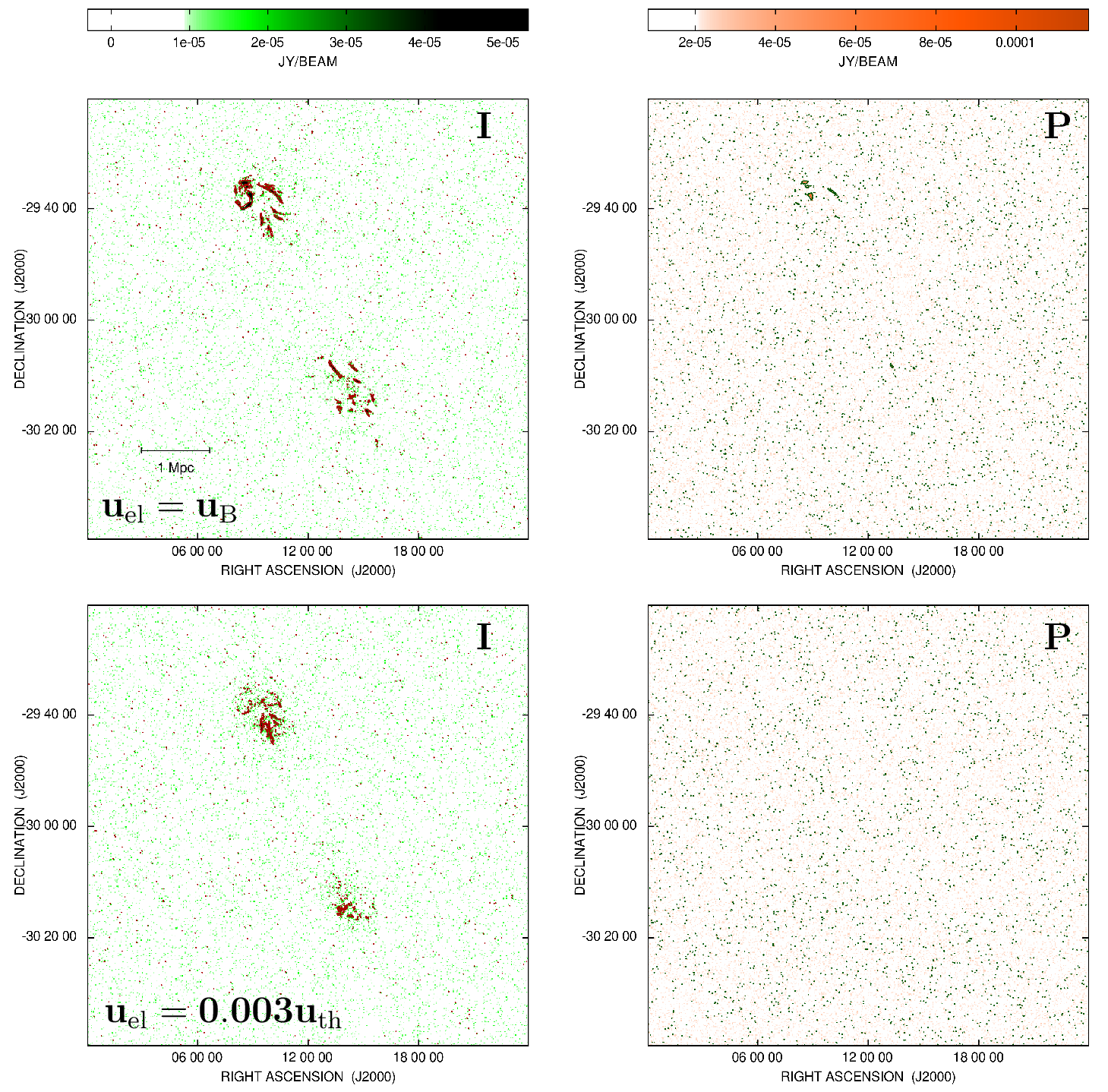}
    \caption{Expectations from a polarization survey with SKA MID AA4 telescopes in Band\,2 (0.95-1.67\,GHz), considering an observing time of 15\,min, a spatial resolution of 2$^{\prime\prime}$, and a channel width of 1\,MHz (see text for further details). On the left total intensity emission is shown, on the right polarized intensity corresponding to the peak in the Faraday depth spectrum. Top panels show expectations for an equipartition scenario, while bottom panels for a scenario with energy density in relativistic electrons proportional to that in thermal particles. A size of 1\,Mpc is shown in the top left panel for comparison.}
    \label{fig:skamid_15min}
\end{figure}
In Fig.\,\ref{fig:skamid_15min}, we present the results for this observing setup. Left panels show total intensity images, while right panels the polarisation images corresponding
to the peak in the Faraday depth spectrum along each line of sight in the cube, after correcting for the polarisation bias according
to \cite{George2012}. Top row refers to the equipartition scenario, while the bottom row to a relativistic electron energy density proportional to the thermal one. Contours are drawn at the corresponding 3\,$\sigma$ for all images. With this observing setup, total intensity images reveal only the brightest regions of the diffuse emission at the centre of the two clusters. Polarized emission is almost completely buried in the noise.

\begin{figure}[h]
    \centering
	\includegraphics[width=1\columnwidth]{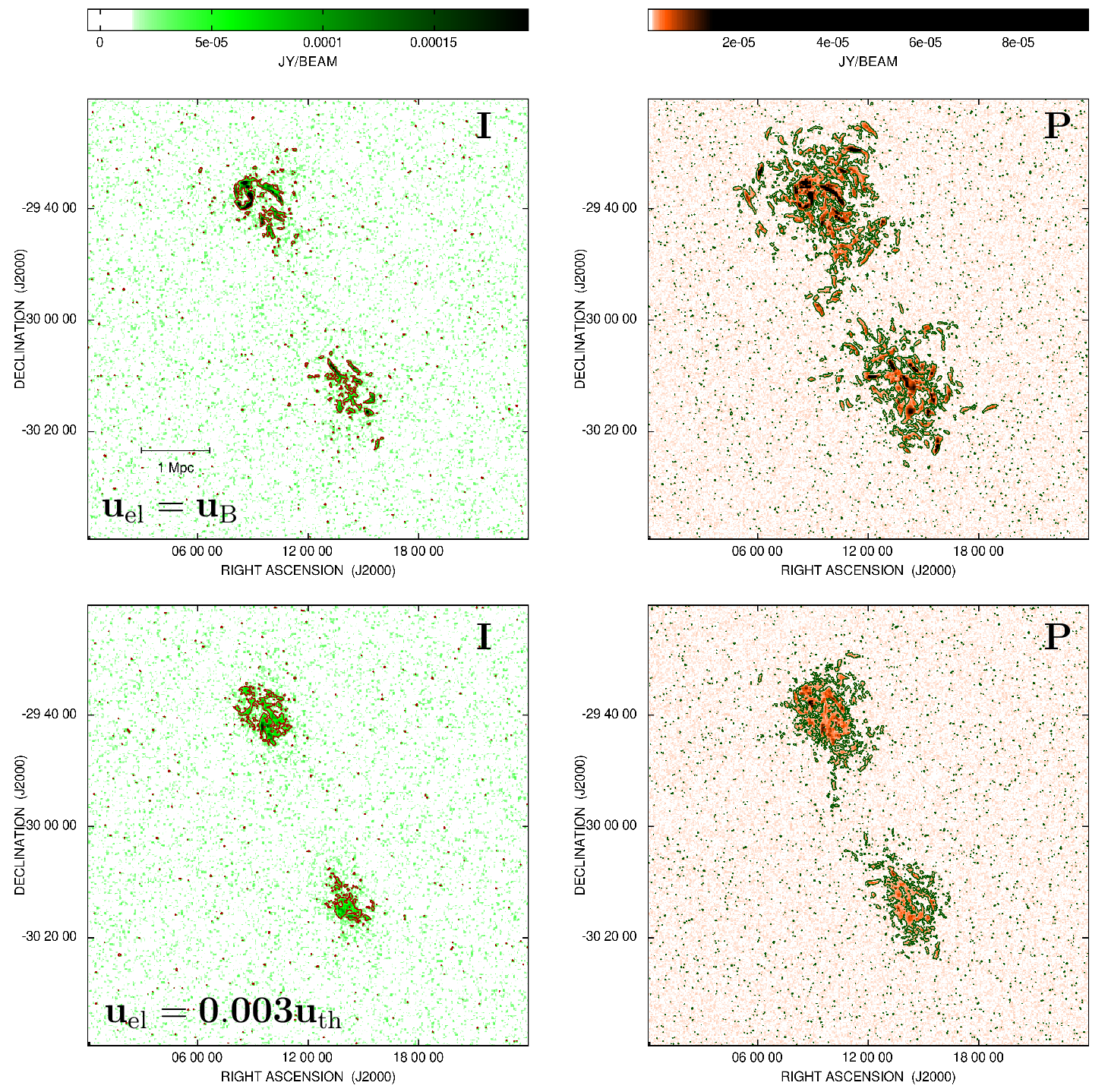}
    \caption{Expectations from the polarization survey with SKA MID AA4 telescopes, considering an observing time of 50\,h and a spatial resolution of 17$^{\prime\prime}$. On the left total intensity emission is shown, on the right polarized intensity corresponding to the peak in the Faraday depth spectrum. Top panels show expectations for an equipartition scenario, while bottom panels for a scenario with the energy density in relativistic electrons being proportional to that in thermal particles. A size of 1\,Mpc is shown in the top left panel for comparison.}
    \label{fig:skamid_50h}
\end{figure}
To enhance the diffuse synchrotron emission, which is characterized by low-brightness levels, deep and low-resolution observations are more suitable.
At the same time, high spatial resolution is needed in order to discriminate between diffuse emission and possible embedded discrete sources and in order to study filamentary features if present. Convolution of high-spatial resolution images after subtraction of the discrete sources could help, but addressing the impact of point source subtraction on the final noise of the image would be necessary. Therefore, a good compromise is represented by imaging at a spatial resolution between 15$^{\prime\prime}$ and 20$^{\prime\prime}$. This spatial resolution is more suitable to detect diffuse extended emission, making still possible to discriminate between diffuse emission and embedded discrete sources and investigate possible filamentary structures. Therefore, we produced additional synthetic images at 17$^{\prime\prime}$ of spatial resolution and an observing time of 50\,h. The frequency range and spectral resolution have not changed. The sensitivity that can be reached with this observing setup is 12.6\,$\mu$Jy/beam in total intensity and 0.24\,$\mu$Jy/beam in Stokes Q and U.
In Fig.\,\ref{fig:skamid_50h}, we present the resulting images. Left and right panels, top and bottom rows, and contours are as described for Fig.\,\ref{fig:skamid_15min}. With these spatial resolution and observing time, a better detection of the total intensity emission is possible with respect to  Fig.\,\ref{fig:skamid_15min}, even if only in the cluster inner regions, because confusion noise dominates, see Table\,\ref{tab:sigma}.
On the contrary, polarization images are not dominated by confusion noise and this allows us a better imaging of the emission. The assumption of equipartition results into a slightly higher brightness with respect to the other scenario. Therefore, in this case, the emission in the periphery of the two galaxy clusters is better recovered. The polarized emission associated with the filament of gas connecting the two clusters also starts to emerge. In the case of relativistic electron energy density proportional to the thermal one, the brightness is fainter and, as a consequence, we have a good mapping of the polarized emission in the central regions of the two clusters, while in the periphery and along the filament connecting the two clusters only the brightest patches can be detected.

Overall, the sensitivity values adopted in this work are summarized in Table\,\ref{tab:sigma}. Since confusion noise in the Stokes Q and U, $\sigma_{\rm conf, Q,U}$, is not included in the present version of the SKAO Sensitivity Calculator and it might affect deep low-spatial resolution data, we estimated it following \cite{Loi2019a} and verified it is negligible at the selected spatial resolutions and frequency.

\begin{table}[h]
	\centering
	\caption{Full bandwidth sensitivity and confusion noise for Stokes parameters I, Q and U for the observing setup considered in this work.}
	\label{tab:sigma}
	\begin{tabular}{cccccc} 
		\hline
        Observing time &Resolution&$\sigma_{\rm I}$ &$\sigma_{\rm conf, I}$& $\sigma_{\rm Q,U}$&$\sigma_{\rm conf, Q,U}$\\
        &$\prime\prime$&$\mu$Jy/beam& $\mu$Jy/beam&$\mu$Jy/beam& $\mu$Jy/beam\\
        \hline
        15\,min&2&6.0&0.13&6.0&0.0004\\
        50\,h&17&12.6&12.6&0.24&0.03\\

\hline
	\end{tabular}
\end{table}

\section{Discussion}

Our results demonstrate that polarization observations are not affected by confusion noise that, on the contrary, dominates total intensity imaging, see Table\,\ref{tab:sigma}. As a consequence, polarized emission can be detected even if total emission is below the noise level. Images expected from an SKA-Mid Band\,2 polarization survey with AA4 telescopes at 2$^{\prime\prime}$ do not allow us to properly image diffuse synchrotron emission from the centre of galaxy clusters to filaments connecting them, neither in total intensity nor in polarization, because they are not sensitive enough. To this end deeper observations are necessary. For an equipartition scenario, we show that an observing time of 50\,h and a spatial resolution of 17$^{\prime\prime}$ are necessary in order to map diffuse synchrotron emission in polarization both at the centre of galaxy clusters and in their periphery as well as along filaments connecting them. If the relativistic energy density is proportional to the thermal one, instead, with this observing setup, a proper recovery of the polarized emission is possible only in the cluster central regions, because of the resulting lower brightness level. In this case, observations deep about six hundred hours are necessary to detect and image polarized emission from the cluster periphery and beyond. The detection of the polarized emission with deep pointed observations is possible due to a better sensitivity compared to survey data. Moreover, the systems considered here are significantly disturbed due to the ongoing merger. Cluster merger events are powerful phenomena expected to release a significant amount of energy in the ICM.
This energy is injected at large spatial scales and then turbulent cascades may be generated \citep[see, e.g.,][]{RoettigerandBurns1999}. Consequently, during the initial phase, we can expect magnetic fields ordered on large scales. This implies a scarce depolarization of the signal. Indeed, as shown in Fig.\,14 by \cite{Vacca2024}, the magnetic field fluctuates on scales from 100 to 300\,kpc. 
Since the angular resolution considered here, i.e. 17$^{\prime\prime}$ corresponds to a linear scale of 20-25\,kpc, being at least a factor five smaller than the fluctuation scale of the magnetic field, and the distribution of the gas density is smooth (see top left panel in Fig.\,1 by \citealt{Vacca2024}), we expect a negligible beam depolarization.
We note that, given the size of the field considered here, these results can be obtained  with the observing time computed with the sensitivity calculator and stated above, if techniques to correct off-axis leakage will be developed and implemented. Otherwise, an observing time a factor two larger will be necessary in order to cover the full field of view with two pointings and ensure proper polarization measurements.
 
\subsubsection*{Magnetic field characterization and link with thermal and dynamical properties of the system}
As described in the previous section, during cluster mergers magnetic fields are expected to show a significant degree of order as a consequence of the energy injection on large spatial scales. In order to verify if these expectations are met, a detailed assessment of the magnetic field properties is necessary. Polarization images of diffuse synchrotron sources can be exploited in order to characterize the magnetic field strength and structure in combination with total intensity images, if available, or alone. When detected, diffuse synchrotron sources provide a good sampling
of the cluster area and, therefore, can play a key role in complementing the information from grids of Faraday rotation measurements towards background discrete radio sources \citep[see, e.g.,][]{Loi2019b}. 
To date a detailed characterization of magnetic field strength and structure is available only for about fifteen systems. By using the results from these studies, here we investigate a possible connection between the thermal and dynamical properties of the system and the intracluster magnetic field power spectrum. 
We exploit information on central thermal gas density, central strength of the intracluster magnetic field and its auto-correlation length, that encloses information about the magnetic field power spectrum\footnote{Following \cite{Ensslin2003}, we define the magnetic field auto-correlation length as 
\begin{equation}
\Lambda_{\rm B}=\frac{3\pi}{2}\frac{\int_{k_{\rm min}}^{k_{\rm max}}|B_{k}|^2k\mathrm{d}k}{\int_{k_{\rm min}}^{k_{\rm max}}|B_{k}|^2k^2\mathrm{d}k}
    \end{equation}
where $k = \frac{2\pi}{\Lambda}$ is the wave number, and $k_{\rm min}$ and $k_{\rm max}$ are the minimum and maximum wave number of the magnetic field power spectrum $|B_{k}|^2$.}. Table\,\ref{tab:data} summarizes all data used in the following\footnote{Please note that for Hydra and 3C31 we converted the $B_{\rm rms}$ values by \cite{Laing2008} into $\langle B_0 \rangle$. With respect to 3C31, we adopted the magnetic field estimate without including X-ray cavities.}\textsuperscript{,}\footnote{For each galaxy cluster, we also display the reference for the intracluster magnetic field parameters. We refer to these papers for the reference of the central thermal gas density values.} 
that have been corrected for the cosmology\footnote{Magnetic field values have been corrected for the cosmology assuming $\langle B \rangle \propto n^{\eta}$, where the value of $\eta$ has been taken from the paper cited in Table\,\ref{tab:data} for each galaxy cluster.}. 
Uncertainties were not always reported in literature, therefore, we adopt a relative uncertainty of 35\,percent on the magnetic field strength and 40\,percent on the length-scale, that have been estimated by using available observed relative uncertainties. We compare these data with properties of simulated systems presented in this work.

\begin{figure}[h!]
    \centering
	\includegraphics[width=1\columnwidth]{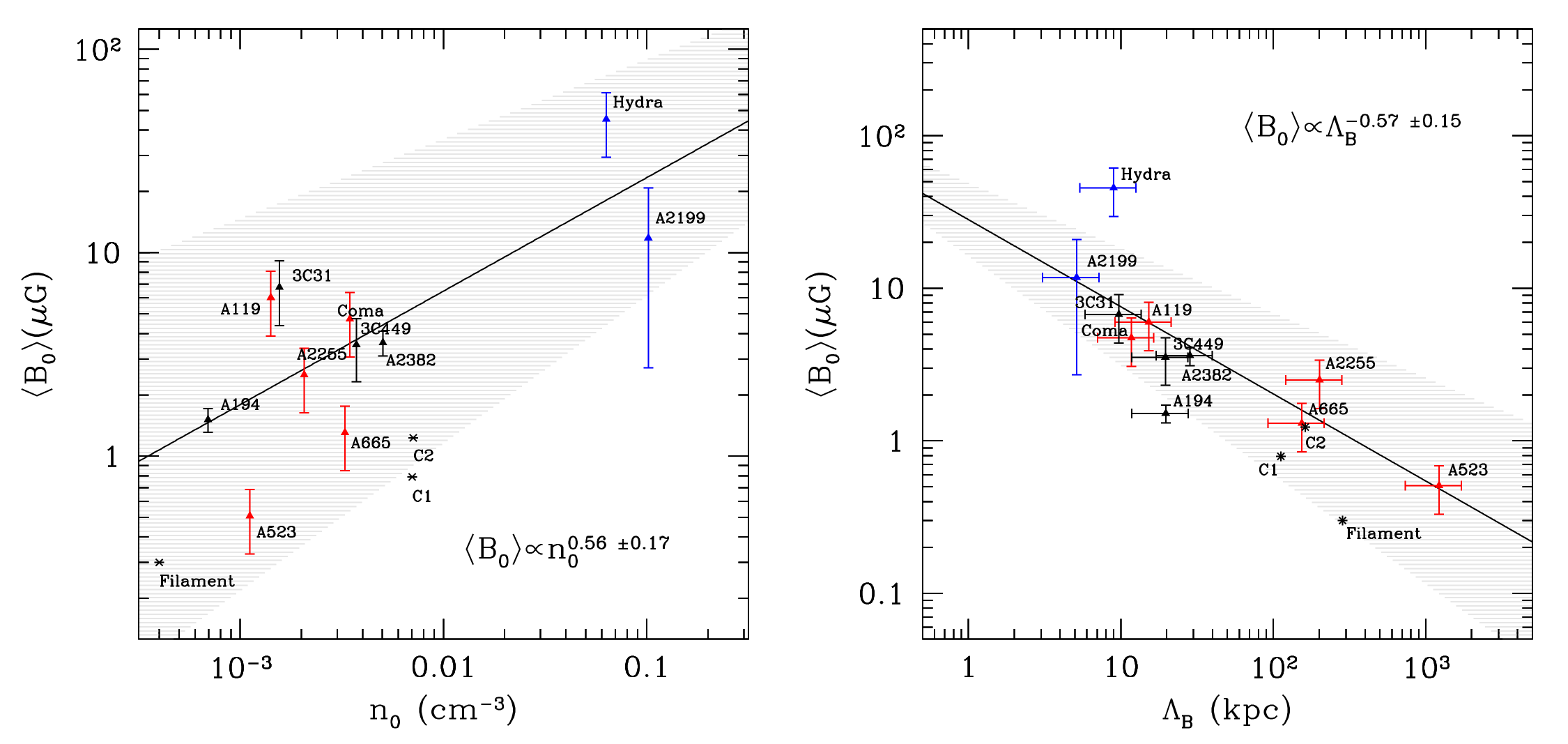}
    \caption{Left panel: central intracluster magnetic field strength versus central thermal gas density. Right panel: central intracluster magnetic field strength versus intracluster magnetic field auto-correlation length. 
    Cool core galaxy clusters are shown in blue, merging cluster in red, and intermediate clusters or groups in black. 
    Simulated systems are plotted as stars. References for observed data points are given in Table\,\ref{tab:data}, while for simulations we refer to \cite{Vacca2024}.}
    \label{fig:B0_vs_Lb}
\end{figure}

In Fig.\,\ref{fig:B0_vs_Lb}, we show the central intracluster magnetic field strength versus the central thermal gas density in the left panel  and versus its auto-correlation length in the right panel. 
Blue dots mark cool core systems, red dots merging galaxy clusters, and black dots galaxy groups or galaxy clusters without major mergers in place. 
\begin{table}[h]
	\centering
	\caption{The table summarizes the central magnetic field strength, magnetic field auto-correlation length, and central thermal gas density values used in this work after conversion for the cosmology. The dynamical state of each system (M = merger, CC = cool core, and NCC = non cool core) and the reference for the magnetic field power spectrum determination are also given.}
	\label{tab:data}
	\begin{tabular}{llllllll} 
		\hline

        Name &$\Lambda_{\rm B}$& $\langle B_{0} \rangle$ &$n_0$ &State& Radio reference\\
             & kpc             & $\mu$G            &     0.001\,cm$^{-3}$&&\\
        \hline 
        &&Observations&&\\
        \hline 
A194   &19.72 $\pm$7.89      &1.51$\pm$ 0.20    & 0.69&NCC&\cite{Govoni2017}\\
A119   &15.27 $\pm$6.11      &6.00 $\pm$2.10    &1.42&M&\cite{Murgia2004}\\
A523   &1223.59$\pm$ 489.44  &0.51$\pm$ 0.18    &1.12&M&\cite{Vacca2022}\\
A665   &154.03$\pm$ 61.61    &1.30$\pm$ 0.46    &3.27&M&\cite{Vacca2010}\\
A2199  &5.13 $\pm$2.05       &11.77 $\pm$9.06   &101.71&CC&\cite{Vacca2012}\\
A2255  &201.17$\pm$ 80.47    &2.51$\pm$ 0.88    &2.06&M&\cite{Govoni2006}\\
A2382  &28.40$\pm$ 11.36      &3.61$\pm$ 0.50    &5.04&NCC&\cite{Guidetti2008}\\
3C31   &9.72$\pm$3.89        &6.75$\pm$ 2.36    &1.56&NCC&\cite{Laing2008}\\
3C449  &19.62 $\pm$7.85      &3.52 $\pm$1.21    &3.73&NCC&\cite{Guidetti2010}\\
Coma   &11.73$\pm$ 4.69      &4.72 $\pm$1.65    &3.46&M&\cite{Bonafede2010}\\
Hydra  &8.97 $\pm$3.59       &45.36 $\pm$15.88  &63.14&CC&\cite{Laing2008}\\

\hline
&&Simulations&&&\\
        \hline 
C1        &112.0             &0.79        &   7.01   &M&\cite{Vacca2024}\\
C2        &162.0             &1.23        &   7.10   &M&\cite{Vacca2024}\\
Filament  &285.0             &0.30        &   0.40   &M&\cite{Vacca2024}\\
		\hline
	\end{tabular}
\end{table}
The plot of the central magnetic field strength versus the central gas density reveals a clear separation between cool core galaxy clusters characterized by higher field strengths and gas densities, and merging and intermediate systems characterized by lower values of these quantities, but no significant difference among the last two classes is observed, as already found by \cite{Govoni2017}. The plot of the central magnetic field strength versus its auto-correlation length shows high strengths and small auto-correlation scales in cool core galaxy clusters, progressively lower field strengths and larger auto-correlation lengths in intermediate systems, and the lowest values of magnetic field strength and the largest auto-correlation scales in merging systems.
In particular, merging galaxy clusters with highly distorted morphologies are characterized by the largest auto-correlation lengths. 
These galaxy clusters are A2255, A523, and A665. The two merging galaxy clusters A119, a radio quiet system,  and Coma, 
characterized by a roundish X-rays morphology and
hosting a roundish unpolarized diffuse synchrotron source at its centre, show intermediate properties more similar to systems where significant disturbances have not been detected. 
Overall, relaxed clusters show high central gas densities, high central magnetic field strengths and small fluctuation scales, while merging and non cool core clusters are characterized by lower magnetic field strength and thermal gas density at their center and auto-correlation scales ranging from intermediate to larger values. 
We performed a linear fit to the data in the two plots in log-log scale and obtained for the plot in the left panel $\langle B_0 \rangle \propto n_0^{0.56\pm0.17}$, in agreement within the uncertainty with  $\langle B_0 \rangle \propto n_0^{0.47}$ found by \cite{Govoni2017}. This trend is expected for a magnetic field energy density scaling as the thermal energy density.
In the plot on the right panel, we obtained $\langle B_0 \rangle \propto \Lambda_{\rm B}^{-0.57\pm0.15}$.

We emphasize that in this chapter, the auto-correlation length is calculated in the same way for different galaxy clusters, regardless of their dynamical state or other properties. Furthermore, to minimize any effects that could distort or influence our results, we limited our analysis to a sample of galaxy clusters in which the magnetic field power spectrum (used to derive the auto-correlation length) is constrained assuming the same modeling.
For most clusters in the sample the magnetic field was characterized using Faraday rotation measurements of cluster/background radio sources.
In general, a two-step approach was adopted:
the slope of the power spectrum and the fluctuation scales were constrained by a preliminary two-dimensional analysis of the Faraday rotation fluctuations and signal depolarization, while the central strength and the radial decay of the magnetic field were determined using three-dimensional numerical simulations in which the field structure was held fixed. The two-dimensional analysis of distributions of the observed rotation measurements yields smaller fluctuation scales in relaxed galaxy clusters and larger fluctuation scales in merging system, without applying any constraints. 
We note that the galaxy clusters A665 and A523 follow the same trend, although, in these cases, the magnetic field properties were derived from the analysis of the diffuse emission of the radio halo.

 Studying the magnetic field auto-correlation length, the strength and, more generally, its power spectrum, it is crucial to understand the field's properties and their relationship to other physical characteristics of the cluster. 
The field's properties are likely deeply linked with/ affected by different and relevant physical mechanisms at work at comparable spatial scales. In particular, we expect that turbulence in galaxy clusters 
can be generated by different phenomena: for example, in disturbed systems the merger can be the main actor, while in relaxed galaxy clusters, AGN jets and/or gas sloshing could play a major role \citep[see e.g. ][]{Vazza2012}. These physical mechanisms act on different scales, larger for disturbed galaxy clusters and shorter for relaxed systems. With respect to the auto-correlation scale, therefore, the plot in the right panel of Figure\,\ref{fig:B0_vs_Lb}, suggests that it is strongly linked to the turbulence phenomena in the cluster. 
Moreover, it shows a smooth transition between the two regimes: the cool-core galaxy cluster A2199 sits close-by the merging and intermediate systems, with larger central magnetic field strength and smaller auto-correlation length, as expected from the fit. This smooth transition can be interpreted as a consequence of the different states of evolution of the turbulence: first, during structure formation processes, the turbulence is injected on large spatial scales, then, it propagates in the cluster environment and, finally, gradually decades on small scales \citep[see, e.g., ][]{RoettigerandBurns1999,Subramanian2006}.
This corresponds to different evolutionary stages of galaxy clusters that merge to produce larger systems and then relax in few Gyrs.

The position of galaxy clusters in this plot could therefore reflect the evolution of turbulence and the current evolutionary state of the cluster. Outliers would then correspond to peculiar galaxy clusters. A good example is represented by Hydra\,A that  sits 2$\sigma$ above the correlation. Indeed, this galaxy cluster shows in X-rays typical properties of a cool-core system. However, a recent work by \cite{Girardi2022} based on optical data demonstrates it is characterized by minor substructures. These peculiarities could be the reason of its displacement with respect the behaviour observed for the remaining clusters in the sample. We note however that only two cool-core galaxy clusters are included in this analysis. A larger statistical sample is needed to understand if Hydra\,A is actually an outlier or if relaxed systems follow a different relation with respect to merging and intermediate systems.

 As a comparison in each plot, marked with stars, we include values for the simulated systems presented in this work: the merging galaxy clusters C1 and C2 and the filament connecting them. The central strength and the auto-correlation length of the magnetic field in the simulated galaxy clusters C1 and C2 and in the filament, as well as their central thermal gas densities, 
have been derived directly from the simulated cubes, without further processing, and are consistent, within the observed scatter, with the
observed values for clusters of galaxies with similar properties.

\section{Conclusions}
Diffuse synchrotron sources represent a powerful tool for constraining magnetic field strength and structure in galaxy clusters and along filaments connecting them. When both total intensity and polarized signal can be detected, an analysis relatively independent on the relativistic electrons, can be conducted. However, at the spatial resolution required to detect their low brightness level, the total intensity emission of these sources is often dominated by confusion of background unresolved radio sources. 

We have shown that low-spatial resolution observations in polarization 
are not affected by confusion noise due to the lower density of polarized radio sources with respect to total intensity ones, and can therefore be used to address the study of magnetic fields in galaxy clusters and filaments of the cosmic web when the total intensity emission is buried in the noise. We demonstrated that a shallow SKAO polarimetric survey with AA4 telescopes in the frequency range 0.95 - 1.67\,GHz, will not allow us to properly image diffuse synchrotron sources nor in total intensity neither in polarization. To this end, we will need to resort to deep ($\geq$50\,h) low spatial resolution pointed observations. 
They will allow us a good mapping of the polarized emission both in galaxy clusters and along filaments connecting them, providing a complementary tool to rotation measure grids in constraining magnetization on large-scales in the Universe. 

Our results indicate that we might also be able to shed light on the distribution and energy content of the relativistic particles that generate the observed diffuse synchrotron emission. 
For the two scenarios considered in this work, indeed, we expect different brightness levels and a different morphology of the radio emission. In the equipartition scenario, we expect filamentary structures in the radio brightness distribution. On the contrary, in case of proportionality between the relativistic and the thermal energy density, an emission fainter, more roundish and smoother foreseen. According to our findings, the sensitivity we will reach with deep pointed observations should be sufficient in order to discriminate among the two.

Finally, we show that current detailed studies on magnetization of galaxy clusters reveal that the intracluster magnetic field is characterized by moderate central strengths ($\approx\,\mu$G) and large auto-correlation lengths ($\gtrsim$100\,kpc) in significantly disturbed systems, progressively less moderate field strengths and decreasing
auto-correlation lengths in galaxy groups or galaxy clusters without major mergers in place, and by high strengths ($\gtrsim$10\,$\mu$G) and short fluctuation scales ($\approx$1-10\,kpc) in relaxed galaxy clusters. These findings are consistent with expectations from simulations and suggest a strong link with the scale of turbulence phenomena in clusters. Future SKAO observations will enable a detailed characterization of magnetic field properties in a larger and larger number of systems and consequently put better statistical constraints on properties of large-scale magnetic fields. 

 Future detailed characterization of magnetic fields in galaxy clusters and filaments of the cosmic web will be crucial to understand cosmic magnetism: the knowledge of the magnetic field strength and structure represents a powerful information to shed light on the origin and evolution of cosmic magnetism. Important insight will also come from the investigation of the correlations of thermal and non-thermal properties in these systems that will provide constraint on the magnetic field radial scaling.

\subsubsection*{Acknowledgements}
We thank the referee for the valuable comments and suggestions which helped improve the manuscript. This work was carried out thanks to the funding of the Regione Autonoma della Sardegna, ai sensi della Legge Regionale 7 agosto 2007, n.7 "Promozione della Ricerca Scientifica e dell'Innovazione Tecnologica in Sardegna".

\bibliographystyle{abbrvnat-maxbibnames4}
\bibliography{chapter} 

\end{document}